\documentstyle[aps,manuscript]{revtex}

\begin{document}

\title {CONNECTION BETWEEN THE LARGEST LYAPUNOV EXPONENT, DENSITY FLUCTUATION AND MULTIFRAGMENTATION
IN EXCITED NUCLEAR SYSTEMS\footnote {Supported by National Natural
Science Foundation of China under Grant Nos. 10175093, 10175089
10235030, 10235020 and Science Foundation of Chinese Nuclear
Industry and Major State Basic Research Development Program under
Contract No. G20000774.
\newline
 E-Mail: lizwux@iris.ciae.ac.cn}}

\author {Yingxun Zhang$^{1}$, Xizhen Wu$^{1,2}$, Zhuxia Li$^{1,2,3}$}

\address { 1) China Institute of Atomic Energy, P. O. Box 275(18), Beijing 102413, P. R. China\\
           2) Nuclear Theory Center of National Laboratory of Heavy Ion Accelerator,\\ Lanzhou 730000,P. R. China\\
           3) Institute of Theoretical Physics, Chinese Academic of Sciences, Beijing 100080, P. R. China }
\maketitle

\begin{abstract}
Within a quantum molecular dynamics model we calculate the largest
Lyapunov exponent (LLE), density fluctuation and mass distribution
of fragments for a series of nuclear systems at different initial
temperatures. It is found that the $LLE$ peaks at the temperature
("critical temperature") where the density fluctuation reaches a
maximal value and the mass distribution of fragments is best
fitted by the Fisher's power law from which the critical exponents
for mass and charge distribution are obtained. The time-dependent
behavior of the LLE and density fluctuation is studied. We find
that the time scale of the density fluctuation is much longer than
the inverse LLE, which indicates that the chaotic motion can be
well developed during the process of fragment formation. The
finite-size effect on "critical temperature" for nuclear systems
ranging from Calcium to superheavy nuclei is also studied.\\

\end{abstract} PACS numbers: 05.70.Fh, 05.45.-a, 05.45.+b\newline
\begin{center}
{\bf 1. INTRODUCTION}
\end{center}

Recent interest in the phase transition in finite nuclear systems
\cite{Bon00,Bor02} and the study of the related dynamical feature
has stimulated the investigation of the, so far, obscure relation
between an anomalous increase of fluctuations at a phase
transition and a rapid increase of chaoticity at the microscopic
level. In this pursuit several papers have been published \cite
{Che01,Bal02,Dor01,Dors01,Iso02,Bal00,Bale00,Gul00,Bal95}. In
\cite{Che01,Bal02,Dor01,Dors01,Iso02,Bal00,Bale00} the studies
were carried out for the excited drops made up of particles
interacting via Lennard-Jones (LJ) potential, whereas the authors
in \cite{Gul00} and \cite{Bal95} used the microcanonical lattice
gas model and Vlasov model,respectively.

A way to characterize the dynamics in the phase transition is to
calculate the largest Lyapunov exponent (LLE), which is a measure
of the sensitivity of the system to initial conditions and also
gives an idea of the velocity at which the system explores the
available phase space. In \cite{Nay95} the LLE has been used to
study the solid-like to liquid-like phase transition in LJ
clusters. In this case the LLE can be understood as an average of
the behavior of the system along an infinite trajectory in phase
space. However in the case of nuclear fragmentation under study, a
nucleus evolves from a highly excited state into a set of nucleons
and clusters, it means that a given trajectory in the phase space
will never come back close to initial state of the system. Hence
such the average over the infinite trajectory will erase the
relevant information of the critical behavior. In order to avoid
this feature we have to calculate the local-in-time LLE over an
ensemble of trajectory whose initial condition is consistent with
the nucleus at a given excitation energy. Because of the absence
of boundary conditions, the ergodic theorem could not be applied.
Therefore, we pay a great attention to study the time scales of
the inverse LLE and density fluctuation. We find that the time
scale of the inverse LLE is much less than that of density
fluctuation. This means that the dynamics during
multifragmentation is chaotic enough, as a result the different
events can sample the whole phase space and the ensemble of
trajectories becomes equivalent to the infinitely long trajectory
of the system. In this way the LLE calculated over an ensemble of
trajectories can carry the full information of multifragmentation.

In this paper we employ the quantum molecular dynamics (QMD) model
to study the critical behavior of real nuclear systems where the
effective nuclear force (Skyrme force) and Coulomb force are
included. With the QMD model we first study the time evolution of
both the LLE and density fluctuation to find their time scales
which are very important in this study. Then we study the
chaoticity characterized by the LLE, the density fluctuation, and
the mass (and charge) distribution of fragments at various initial
temperatures, so that we can obtain the correlation between the
characteristic temperatures for the system reaching a maximal
chaoticity, maximal density fluctuation, and attaining a power law
mass spectra. In order to study finite size effects on "critical
temperature", numerical studies are carried out for nuclear
systems ranging from $^{40}Ca$ to the superheavy nucleus
$^{298}114$. We expect that, through our study, a better
understanding of nuclear multifragmentation and its relation to
the Lyapunov instability are obtained.

This paper is organized as follows. In Sec.$2$ we briefly
introduce the quantum molecular dynamics model (QMD) used in our
numerical calculations. Then the calculated results are shown in
section $3$. In Sec.$ 4$, a possible connection between the LLE
and the fluctuation of nuclear density is discussed preliminary.
Finally, a brief summary is given in Sec.$5$.

\begin{center}
\bigskip {\bf 2. MODEL}
\end{center}

The QMD model is employed in describing the dynamic evolution of
an excited nuclear system, which contains not only some quantum
features but also many-body correlations. Therefore, it has been
widely used \cite{Har89,Aic91,Har98} in modelling intermediate and
high energy heavy-ion collisions, and it successfully provides
much dynamic information about nuclear reactions and
multifragmentation due to its practical approach to studying heavy
nuclear systems. Compared with Anti-symmetrized Molecular
Dynamics\cite{Ono92} and Fermionic Molecular Dynamics\cite{Fel97},
 our model treats the effect of the Pauli principle approximately
\cite {wan02}. The phase space constraint proposed by Papa et.
al.\cite {Pap01} is introduced, that is, the one-body occupation
number in a volume $h^{3}$ of phase space centered at ($\vec{r_{i}}$,
$\vec{p_{i}}$), corresponding to the centroid of the wave pocket of particle
i, should always be not larger than 1. For reader convenience, in this section we
briefly introduce the model. In the model, each nucleon is represented by
a coherent state of a Gaussian wave packet
\begin{equation}
  \psi(\vec{r},t)=\frac{1}{(2\pi\sigma^{2}_{r})^{3/4}}
\exp[-(\vec{r}-\vec{r_{i}})^{2}/4\sigma^{2}_{r}]\exp[i\vec{p_{i}}\cdot\vec{r}/\hbar].
\end{equation}
Where, $\vec{r_{i}}$ and $\vec{p_{i}}$ are the centers of the wave
packet of particle $i$ in the coordinate and momentum space,
respectively. $\sigma_{r}$ represents the spatial spread of the
wave packet. Through a Wigner transformation of the wave
function, the one-body phase space distribution function for
N-distinguishable particles is given by
\begin{equation}
f(\vec{r},\vec{p})=\frac{1}{(\pi\hbar)^3}\sum_{i=1}^{N}
\exp[-(\vec{r}-\vec{r}_{i})^{2}/2\sigma^{2}_{r}-(\vec{p}-
\vec{p}_{i})^{2}2\sigma^{2}_{r}/\hbar^{2}].
\end{equation}
The density and momentum distribution functions of a system read
\begin{equation}
\rho(\vec{r})=\int{f(\vec{r},\vec{p})d\vec{p}}=\sum_{i=1}^{N}\rho_{i}(\vec{r}),
\end{equation}
and
\begin{equation}
g(\vec{p})=\int{f(\vec{r},\vec{p})d\vec{r}}=\sum_{i=1}^{N}g_{i}(\vec{p}),
\end{equation}
respectively, where the sum runs over all particles in the system.
$\rho_{i}(\vec{r})$ and  $g_{i}(\vec{p})$ are the density and
momentum distribution functions of nucleon $i$:

\begin{equation}
\rho_{i}(\vec{r})=\frac{1}{(2\pi\sigma^{2}_{r})^{3/2}}
\exp(-\frac{(\vec{r}-\vec{r_{i}})^{2}}{2\sigma^{2}_{r}}),
\end{equation}
\begin{equation}
g_{i}(\vec{p})=\frac{1}{(2\pi\sigma^{2}_{p})^{3/2}}
\exp(-\frac{(\vec{p}-\vec{p_{i}})^{2}}{2\sigma^{2}_{p}}),
\end{equation}
where $\sigma_{r}$ and $\sigma_{p}$ are the widths of wave packets
in the coordinate and momentum space, respectively, and they
satisfy the minimum uncertainty relation:
$\sigma_{r}\cdot\sigma_{p}=\frac{\hbar}{2}$. The time evolution of
$\vec{r_{i}}$
 and $\vec{p_{i}}$ is governed by Hamiltonian equations of
motion
\begin{equation}
\dot{\vec{r_{i}}}=\frac{\partial{H}}{\partial{\vec{p_{i}}}},
\dot{\vec{p_{i}}}=-\frac{\partial{H}}{\partial{\vec{r_{i}}}}.
\end{equation}
The Hamiltonian $H$ is made up of the kinetic energy and the
effective interaction potential energy:
\begin{equation}
H=E_{k}+U,
\end{equation}
where
\begin{equation}
E_{k}=\sum_{i=1}^{N}\frac{\vec{p_{i}}^{2}}{2m}.
\end{equation}
The effective interaction potential energy reads
\begin{eqnarray}
U=\frac{\alpha}{2}\sum_{i=1}^{N}<\frac{\rho}{\rho_{0}}>_{i}
+\frac{\beta}{3}\sum_{i=1}^{N}<\frac{\rho^{2}}{\rho_{0}^{2}}>_{i}
+\frac{C_{s}}{2}\int{\frac{(\rho_{p}-\rho_{n})^2}{\rho_{0}}d\vec{r}}\\
\nonumber +\frac{C_{y}}{2}\sum_{i=1}^{N}\int{\rho_{i}(\vec{r})
\frac{\exp[-\gamma|\vec{r}-\vec{r\acute{}}|]}
{|\vec{r}-\vec{r\acute{}}|}\rho_{j}(\vec{r})d\vec{r}d\vec{r\acute{}}}\\
\nonumber +\frac{1}{2}\sum_{i\neq{j}(i,j for protons)
}\int{\rho_{i}(\vec{r})\frac{e^{2}}
{|\vec{r}-\vec{r'}|}\rho_{j}(\vec{r})d\vec{r}d\vec{r'}}.
\end{eqnarray}

The parameters $\alpha,\beta,C_{s}$ and $C_{y}$ in the model are
taken to be the same as in Ref\cite{Fan99}. In order to prepare a
ground state nucleus, we first calculate the neutron and proton
density distribution, binding energy and nuclear radius of the
ground state by the relativistic mean field theory (RMF)
\cite{Mao99}. Then the position of each nucleon in the nucleus is
sampled according to the density distribution obtained. The
momentum of each wave packet is assigned randomly between zero and
local Fermi momentum $P_{f}(r)$ obtained  by the local density
approximation. Each created nucleus is examined according to the
properties of the ground state ( i.e., the binding energy and the
nuclear radius) and the time evolution of the binding energy and
root-mean-square radius. Only those whose properties are in
consistent with those of the ground state are accepted as initial
ground state nuclei (see, Ref.\cite{wan02} for detail). The
initial excited nuclei are obtained by the following procedures:
the position of each nucleon is taken to be the same as its ground
state, but the momentum of each nucleon is re-sampled according to
the Fermi-Dirac distribution at a certain value of chemical
potential and temperature T. By varying the temperature, we put
different initial kinetic energies (initial excitation energies)
into  nuclear systems. With this procedure, we prepare a series of
initial excited nuclei, such as $^{40}Ca$, $ ^{58}Ni$, $^{90}Zr$,
 $^{124}Sn$, $^{144}Nd$, $^{197}Au$, $^{208}Pb$, $^{266}Ra$,
 $^{238}U$, and $^{298}114$ for dynamic studies.

\begin{center}
\bigskip {\bf 3. NUMERICAL STUDY OF MULTIFRAGMENTATION}
\end{center}

\bigskip {\bf 3.1 LARGEST LYAPUNOV EXPONENT}

The largest Lyapunov exponent is defined
as\cite{Nay95,Bon00,Bal95}
\begin{equation}
\lambda=\lim_{n\rightarrow\infty}\frac{1}{n\tau}
\ln\frac{\|d\vec{X_{n}}\|}{\|d\vec{X_{0}}\|}.
\end{equation}
The quantity $\|d\vec{X_{n}}\|$ is the phase space distance
between two trajectories corresponding to two concerned events at
time $t=n\tau$ and the phase space distance $\|d\vec{X}\|$ reads
\begin{equation}
\|d\vec{X}\|=\sqrt{\sum_{i=1}^{N}[(\vec{r_{1}}/rms-\vec{r_{2}}/rms)^{2}
+(\vec{p_{1}}/avp-\vec{p_{2}}/avp)^{2}]_{i}}.
\end{equation}
Where the sum runs over all $N$ nucleons of the system, the
subscripts 1 and 2 refer to two events which differ of an
infinitesimal quantity of $\|d\vec{X_{0}}\|=10^{-7}$ or less at
initial time. In Eq.(12), the dimensionless coordinate and
momentum, scaled by the root-mean-square radius ($rms$) and the
average momentum ($avp$), respectively, are used. In our model,
the trajectory $\vec{X}(t)$ is a function of a set of
$\{{\vec{r_{i}}, \vec{p_{i}}}\}$ which defines the states of a
nucleus.

 In numerical calculations of the LLE, the initial
excited nuclei are created by the method mentioned in section 2. For each
temperature $T$, $50$ test events are generated, and for each test
event 40 other different events are generated, each of them
differing from the corresponding test event by $\|d\vec{X_{0}}\|$
at the initial time. The LLE is obtained by averaging over
trajectories of all events evolving according to a set of
equations of motion (Eq.(7)).
 To show the feature of time evolution of the LLE
at different temperatures, as an example, we plot $\lambda$(t) for
$^{208}Pb$ at temperatures of $T=2$, $11$, $30 MeV$ in Fig.1. From
this figure one can see that the behavior of $\lambda$(t) for T=11
MeV is quite different from the cases of T=2 and 30 MeV. For the
case of $T=2$ MeV the $\lambda$(t) reaches a constant value after
$T=110$ fm/c, and for T=$30 MeV$ the $\lambda$(t) firstly
decreases with time and finally approaches to an asymptotic value.
Whereas for the case of T=11 MeV there appears a plateau in the
LLE from time 130 fm/c to 175 fm/c and after then the LLE
decreases again. The saturation behavior in T=2 and 30 MeV is
simply  due to the fact that the available  phase space is
limited. Whereas for the case of T=11 MeV the LLE  maintains at a
constant value only during the finite time in which
multifragmentation takes place. After fragment formation the
collective expansion motion of the fragmenting system may play a
major role, this ordered motion results in a reduction of the LLE.
Considering this feature, we take the $\lambda$(t) value at the
plateau as the LLE of the fragmenting system. In order to find a
"critical temperature" we calculate the LLE of selected systems at
different temperatures in step of 1 MeV, and as examples, we show
the results for $^{124}Sn$ and $^{208}Pb$ in Fig. 2.  One can find
that, with the increase of temperature T, the $LLE$ increases
until it reaches a maximal value at a certain temperature, and
afterwards the LLE decreases as temperature further increases. We
call the temperature corresponding to the maximal LLE as the
"critical temperature". This temperature is just the one at which
a plateau appears in the time evolution of $\lambda$(t). The
behavior of the LLE shown in Fig. 2 is also found in all nuclear
systems studied (for example, $^{144}Nd$, $^{197}Au$, $^{226}Ra$,
$^{238}U$, $^{298}114$, $^{40}Ca$, and $^{58}Ni$)and was observed
in Ref. \cite{Dor01}. Here we notice that the numerical evaluation
about the "critical temperature" is a bit uncertain because the
saddle-point has to be found out within a fluctuating signal. This
uncertainty is even a bit more for light nuclear systems, due to
stronger finite size effects. However, for heavy nuclear systems,
since the observed change of the signature is strong enough, the
peak in the LLE is well pronounced. The behavior of the LLE as a
function of temperature can be easily understood as follows: The
raising branch is obviously due to the increase of fluctuation
with temperature, and the presence of the maximum in the LLE,
which signals a transition from a chaotic to a more ordered
motion, results form multifragmention; The behavior of the descent
branch of $\lambda$$\sim$T can be traced to the fact that in this
temperature region the system breaks up very soon and the ordered
expansion collective motion dominates the evolution of the system.

The mass dependence of the "critical temperature" for nuclei
ranging from $^{40}Ca$ to $^{298}114$ is shown by the line
with solid squares in Fig.3. One can see from the curve that: as
the system size increases, the "critical temperature" increases
slightly, and for the systems heavier than $^{197}Au$, the
"critical temperature" approximately approaches to a constant
value of about $T_{c}=11MeV$. The trend of the dependence of the
"critical temperature" for finite systems on the nuclear mass is
consistent with the other model calculations \cite{Jaq84,Ell00,Nat02}.

\bigskip {\bf 3.2  DENSITY FLUCTUATION }

In the QMD model, the many body correlation can be taken into account,
which allows us to study the density fluctuation defined as
\begin{equation}
\sigma_{\rho}^{2}=\frac{<\rho^{2}(t)>-<\rho(t)>^{2}}{<\rho(t)>^2}.
\end{equation}
Here,
\begin{equation}
<\rho(t)>=\int{\rho(t)\rho(t)d\vec{r}},
\end{equation}
and
\begin{equation}
<\rho(t)^{2}>=\int{\rho(t)^{2}\rho(t)d\vec{r}}.
\end{equation}
The integration is over the whole space. Since the system studied
is an isolated one, the density fluctuation should asymptotically
approach to a saturation value under the influence of the mean
field and the kinetic energy term. Fig. 4 shows the time evolution
of the density fluctuation at temperatures of $T=3$, $11$ and $20$
MeV for $^{208}Pb$. From the figure one sees that for the case of
T=3 MeV, the $\sigma_{\rho}^{2}$ is very small and there is a very
slow increase in the density fluctuation. This is because of the
effect of neutron evaporation. For T=20 MeV, the density
fluctuation sooner reaches a saturation value since the system
breaks up very soon. While for the case of T=11 MeV, the density
fluctuation grows abnormally till t=175 fm/c and after then there
appear small jumps. The abnormal growth rate and jumps seen in the
curve signal the process of fragmentation. In order to see the
saturation behavior the effect resulting from evaporation process
should be subtracted. We show the density fluctuation with the
neutron evaporation subtracted for T=3 MeV case and the secondary
evaporation subtracted for T=11 MeV case in the small figure
inserted in Fig.4. For T=20 MeV case the system breaks up very
fast and the evaporation almost has no effect on the density
fluctuation of the system. From the inserted figure, one sees the
saturation behavior of $\sigma_{\rho}^{2}$ for T=3 MeV case and
the asymptotically saturation behavior of $\sigma_{\rho}^{2}$ for
T=11 MeV case, respectively. By comparing the time dependent
behavior of the density fluctuation with that of the LLE, we find
quite different time scales: the time scale for density
fluctuation growth ($\sim$ 150fm/c) is much longer than the
inverse largest Lyapunov exponent ($\sim$ 40 fm/c). This finding
indicates the fact that the dynamics during fragmentation of the
nuclear system is chaotic enough. Based on this finding we are
allowed to use the Lyapunov exponent to characterize the dynamics
of fragmentation in finite nuclear systems (see introduction).

 Then we study the evolution of saturation values of
$\sigma^{2}_{\rho}$ with temperatures for the systems studied in
Sec. 3.1. We find that the behavior of $\sigma^{2}_{\rho}$ $\sim$
T is quite similar with that of the LLE$ \sim$ T ( shown in Fig.
2). Similarly, we can extract the "critical temperature" from the
maximal value of $\sigma^{2}_{\rho}$ $\sim$ T. The mass dependence of
the "critical temperature" extracted by the density fluctuation is
also shown in Fig. 3.  We find that
the "critical temperatures" obtained from both the density
fluctuation and the largest Lyapunov exponent are in well
coincidence. It implies that the abnormal fluctuation in the
density emerges from the deterministic chaos and thus a small
uncertainty in the initial condition can produce a large dynamical
fluctuation in final observables.

\bigskip {\bf 3.3 MULTIFRAGMENTATION}

In this section we study the mass and charge distributions of
fragments for the selected systems mentioned in section 3.1 at
different temperatures. In our calculation, the fragment
recognition is in terms of the conventional coalescence
model\cite{Kru85}, in which particles with relative momenta
smaller than $P_{0}$ and relative distance smaller than $R_{0}$
are considered to belong to one cluster. Here $R_{0}$ and $P_{0}$
are taken to be 3.5 fm and 300 MeV/c, respectively
\cite{Zha99,Li02}. In Fig. 5 we show the calculated mass
distributions at different temperatures for systems of $^{124}Sn$
and $^{208}Pb$. From this figure we can see that at low
temperatures ( for example, T=4 MeV ) only a few nucleons are
vaporized, and the mass of residues is peaked near the original
nucleus (U shaped mass spectra). However, for high temperatures (
for instance, T=25 MeV ) the system breaks up into nucleons and
light fragments, thus the mass distribution of fragments peaks at
the very small mass number, and the system is considered to be at
a "vapor" phase. If the initial temperature is in between, for an
example, at T= 8 MeV the system starts to fragment and the
coexistence of "liquid" and "vapor" may appear. At T=10 MeV for
$^{124}Sn$ and 11 MeV for $^{208}Pb$, the mass of fragments is
distributed over a wide range from free nucleon to about a half of
the total mass of the system, and the maximal fragmentation seems
to appear. The sequence of shapes of above mass spectra is the
same as the one predicted by Fisher's model\cite{Fis67} of
liquid-gas phase transition. In the latter model the probability
of having a drop of size A in the vapor is given by
\begin{equation}
P(A)=Y_{0}A^{-\tau}exp[-(\mu_{l}-\mu_{g})A+4\pi*r_{0}^{2}\sigma(T)A^{2/3}].
\end{equation}
Here, $\mu_{l}$ and $\mu_{g}$ are the chemical potential of the
liquid and vapor phases, and $\sigma$ is the surface tension. In
the critical point, $\mu_{l}$ equals $\mu_{g}$ and $\sigma(T_{c})$
equals zero, then the power law is obtained. In order to fit the
power law, we give a double-logarithmic plot of the mass( and
charge) spectra at T =11 MeV for $^{208}Pb$ in Fig. 6. The solid
squares (open circles) denote the numerical results for mass (charge)
spectra, and the dashed lines denote their best fits to mass and
charge spectra with $\chi^{2}$ of 0.678 and 1.140, respectively.
From the best fits to mass and charge spectra for all selected
systems we can obtain the critical temperature $T_{c}$ and the
critical exponents $\tau _{m}$( for mass) and $\tau _{z}$ (for
charge). Table 1 lists those results with the values of
$\chi^{2}$. We note that all extracted exponents are larger than
2.0, and they are in agreement with what is expected from the
Fisher's power law for the nuclear liquid-gas phase transition.
The calculated critical exponents for charge distributions are
quite close to the recently obtained experiment value of
$\tau_{z}\sim 2.35\pm0.05$ \cite{Buk02}. We find that the critical
temperatures obtained by the best fit to power law are the same as
those obtained by means of the $LLE$ and density fluctuation. This
means that the power law behavior of mass spectra is closely
related to the pronounced peak in the LLE and density fluctuation.
As is shown in section 3.1 that the maximum in the LLE corresponds
to the largest local rate of the divergence of trajectories, thus
to the maximum in the available phase space of trajectories. In
fact, if the dynamics is chaotic, strong fluctuation is expected
from one microscopic ( collision) event to another, each event
ending in a different region of the available phase space. On the
average, therefore, if chaoticity is strong enough, the population
of the final channels will be dominated by the available phase
space. From this point of view, Fig. 5 may tell us that a very
large population of the decay channels can be obtained even at the
"critical temperature" $T_{c}$ at which the possible
configurations show the maximal variation. Thus, the
correspondence between multifragmentation and maximum chaoticity
in microscopic level can been established.

\begin{center}
\bigskip {\bf 4.BRIEF DISCUSSION OF CONNECTION BETWEEN THE LLE AND DENSITY
         FLUCTUATION}
\end{center}
In this part, we further discuss the connection between the
Lyapunov exponent of trajectories around an ensemble of initial
states consistent with given initial heated nuclei and the
fluctuation of nuclear density related to the decomposition of a
nucleus into fragments. As is described in section 3.1 that the
trajectory $\vec{X}(t)$ is a function of a set of $\{{\vec{r_{i}},
\vec{p_{i}}}\}$ which defines the states of a nucleus. The phase
density $\varrho(\vec{X})$ is, therefore, also a function of the
set of $\{{\vec{r_{i}}, \vec{p_{i}}}\}$.

The general connection between the LLE and the phase density
$\varrho(\vec{X})$ is given in Refs. \cite{Eck85,Gu90} as
\begin{equation}
\lim_{t\rightarrow{\infty}}{\frac{1}{2t}\ln{h(\varrho)}}
=\max_{\vec{X}\in\Lambda}{\lambda(\vec{X})},
\end{equation}
where the heterogeneity of the phase density is defined as
\begin{equation}
h(\varrho)=\int|\frac{\partial\varrho(\vec{X})}{\partial\vec{X}}|^2$d$\vec{X}
  /\int|\varrho(\vec{X})|^2$d$\vec{X},
\end{equation}
where the $\lambda(\vec{X})$ is the LLE of the trajectory
$\vec{X(t)}$ (defined in Eq.(11)), the $\Lambda$ is the non-zero
domain of $|\nabla\varrho^{0}(\vec{X})|$ in phase space and the
$\varrho^{0}(\vec{X})$ denotes the phase density
$\varrho(\vec{X})$ at time t=0. Here the maximum has to be
understood in measure theoretical sense. The expression (17) is
known to be a quite general formula. We will make a qualitative
discussion about
 the connection between the LLE and
density fluctuation by means of expressions of (17) and (18),
although use of expression (17) in the phase transition regime
might be questionable. According to the definition of
$\varrho(\vec{X}(t))$ and $\vec{X}$, the phase density $\varrho$,
in a certain sense, can be roughly considered as a function of
nuclear density $\rho$ (see, Eq.(3)) and momentum distribution $g$
(see,Eq.(4)), thus one can easily deduce that the heterogeneity of
the phase density $h(\varrho)$ increases with
$\langle{\rho^{2}}\rangle-\langle{\rho}\rangle^{2}$, i. e.,
increases with the density fluctuation $\sigma_{\rho}^{2}$. From
expression (17) the LLE of trajectory should increase with the
density fluctuation. Of course, this is only a very qualitative
discussion. However, based on our numerical results the
correspondence of the LLE and the density fluctuation for
different systems can be illustrated. Fig. 7 shows the relation
between the LLE $\lambda(\vec{X})$ and the density fluctuation
$\sigma_{\rho}^{2}$ at different temperatures from 3 MeV to 19 MeV
for systems of $^{124}Sn$ ( Fig. 7(a)) and $^{208}Pb$ ( Fig.
7(b)). One can see that the maximum values of both the LLE and
density fluctuation are located at the same temperature, i.e. the
" critical temperature". There are two branches in
 $\lambda(\vec{X})$ $\sim$ $\sigma_{\rho}^{2}$, one corresponding to the
temperature lower than the "critical temperature" and another
corresponding to the temperature higher than the "critical
temperature". For the low temperature branch, both of the
$\lambda(\vec{X})$ and $\sigma_{\rho}^{2}$ increase as the
temperature increases, whereas for the high temperature branch
they increase as the temperature decreases. Both branches show
that the $\lambda(\vec{X})$ roughly linearly increases with
$\sigma_{\rho}^{2}$. This correspondence between the
$\lambda(\vec{X})$ and $\sigma_{\rho}^{2}$ is qualitatively in
consistence with the discussion based on the expressions of (17)
and (18).

\begin{center}
\bigskip {\bf 5. SUMMARY}
\end{center}
In this work we have systematically studied the fragmentation
process of hot nuclei in terms of the LLE, the density fluctuation
and the mass(charge) spectrum. The character of the LLE at the
"critical temperature" is that not only its value reaches the
largest one, but also there appears a plateau in its time
evolution, which represents the period of formation process of
fragments. Simultaneously, at the "critical temperature" in the
time evolution of the density fluctuation there appear an abnormal
growth rate and jumps which indicate the process of decomposition
of a nucleus into fragments. Our study further demonstrates that
the time scale of the density fluctuation is much longer than the
inverse largest Lyapunov exponent, which means that the chaotic
motion can be well developed during the process of fragment
formation. Therefore, the deterministic chaotic mechanism is
allowed to describe the fragmentation in finite nuclear systems,
and it seems to be of a crucial importance for the phase
transition.

 Our study shows that the $LLE$  peaks at the temperature, at which
 the density fluctuation grows abnormally and the mass distribution of
 fragments fits best to the Fisher's power law, for all selected
 nuclear systems ranging from $^{40}Ca$ to the superheavy
 nucleus $^{298}114$. This means that all three signatures are in coincidence
 at the "critical temperature". As can be seen that the observed changes
 in the key signatures seem to be strong enough and therefore they could
 survive even when associating each point with reasonable error bars caused by
 numerical uncertainties and the approximate treatment of the
 Pauli principle.

We have further investigated finite size effects on "critical
temperature", and it is observed that for systems lighter than
$^{197}Au$ the "critical temperature" increases with mass and for
systems heaver than $^{197}Au$ a saturation value of about
$T_{c}=11MeV$ seems to be reached. \\


\begin{table}[htbp]
\caption{The power law of mass and charge distributions for
systems $^{124}Sn$ ,$^{144}Nd$, $^{197}Au$, $^{208}Pb$,
$^{226}Ra$, $^{238}U$, and $^{298}114$ at their "critical
temperatures". } \vspace{0.3cm}
\begin{tabular}{|c|c|c|c|c|c|c|c|}
\hline
  & $^{124}Sn$ & $^{144}Nd$& $^{197}Au$ &
$^{208}Pb$ & $^{226}Ra$ & $^{238}U$& $^{298}114$  \\
\hline
$T_{c}$  & $10MeV$ &$10MeV$&$11MeV$&$11MeV$&$11MeV$&$11MeV$&$11MeV$ \\
\hline
$\tau_{m}$ & $2.679$  &$2.672$ &$2.696$ &$2.676$ &$2.642$ &$2.700$ & $2.660$\\
$(\chi^{2})$ & $0.510$  &$0.611$ &$0.594$ &$0.678$ &$1.353$ &$0.792$ & $0.531$\\
\hline
$\tau_{z}$ & $2.514$  &$2.496$ &$2.477$ &$2.453$ &$2.406$ &$2.453$ &$2.432$ \\
$(\chi^{2})$ & $0.146$  &$0.256$ &$1.067$ &$1.140$ &$1.184$ &$1.497$ &$1.606$ \\
\hline
\end{tabular}
\end{table}

\begin{figure}
\caption{  The time evolution of
$\frac{1}{t}\ln{\frac{\|d\vec{X(t)\|}}{\|d\vec{X_{0}\|}}}$ for the
system $^{208}Pb$ at temperatures of $T=2$, $11$, and $30MeV$.
        }
\label{fig1}
\end{figure}

\begin{figure}
\caption{  The largest Lyapunov exponent as a function of
temperature for systems of $^{124}Sn$ and $^{208}Pb$.
        }
\label{fig2}
\end{figure}

\begin{figure}
\caption{ The "critical temperature" obtained from the $LLE$ 
 ( line with solid squares) and density fluctuation ( line with open
 circles) for various nuclear systems.
       }
\label{fig3}
\end{figure}

\begin{figure}
\caption{ The time evolution of the density fluctuation for
$^{208}Pb$ at different initial temperatures of $T=3$, $11$, and
$20MeV$.
        }
\label{fig4}
\end{figure}

\begin{figure}
\caption{ The mass distribution of fragments at various
temperatures for systems of $^{124}Sn$ and $^{208}Pb$.
        }
\label{fig5}
\end{figure}

\begin{figure}
\caption{ The double-logarithmic plot of the mass and charge
distribution of fragments at the "critical temperature" for
$^{208}Pb$.
        }
\label{fig6}
\end{figure}

\begin{figure}
\caption{ The relation between the LLE and the density fluctuation
at temperatures from 3 MeV to 19MeV for systems of $^{124}Sn$ (
Fig. 7(a))and $^{208}Pb$ ( Fig. 7( b)).
        }
\label{fig7}
\end{figure}
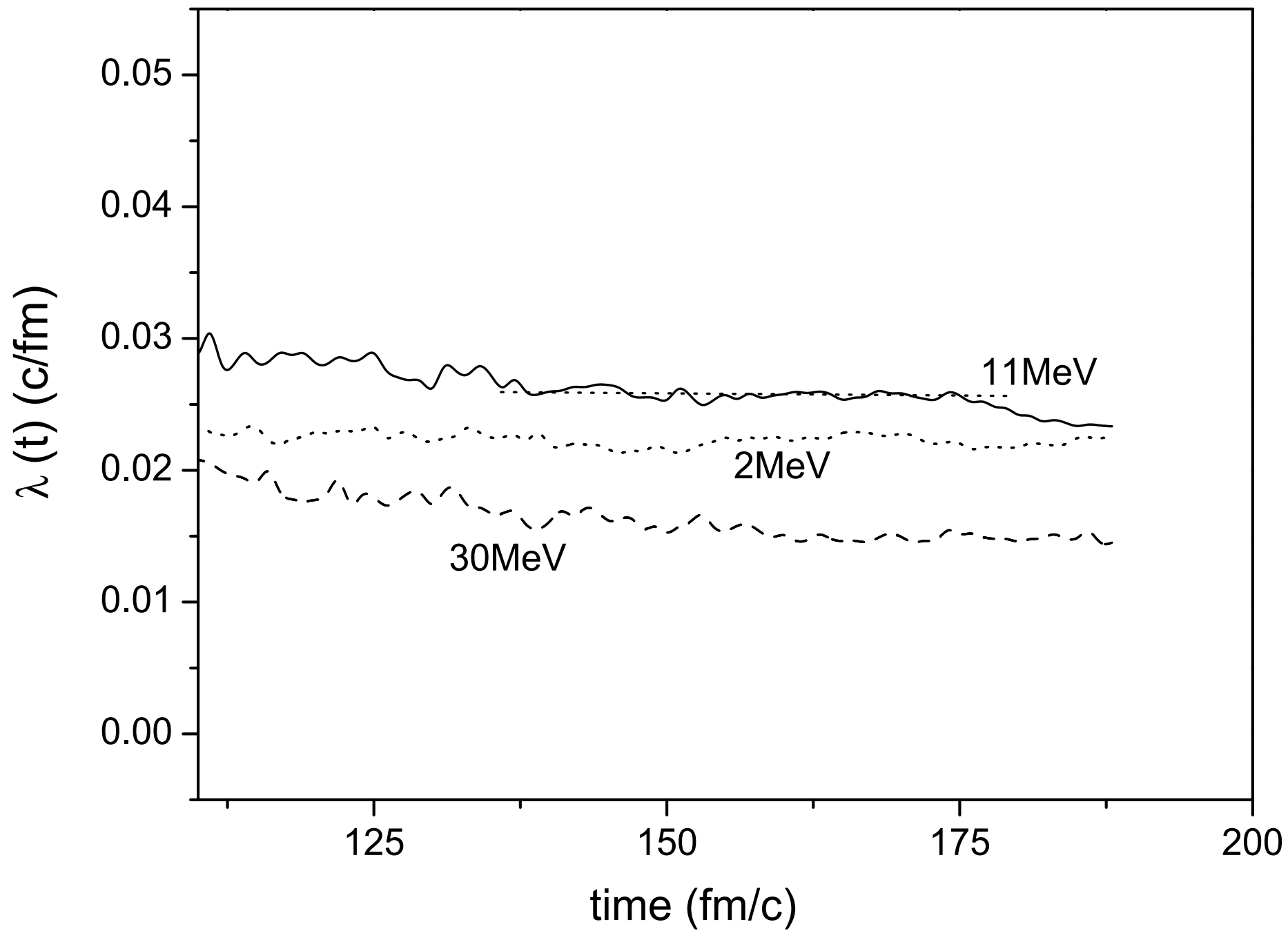
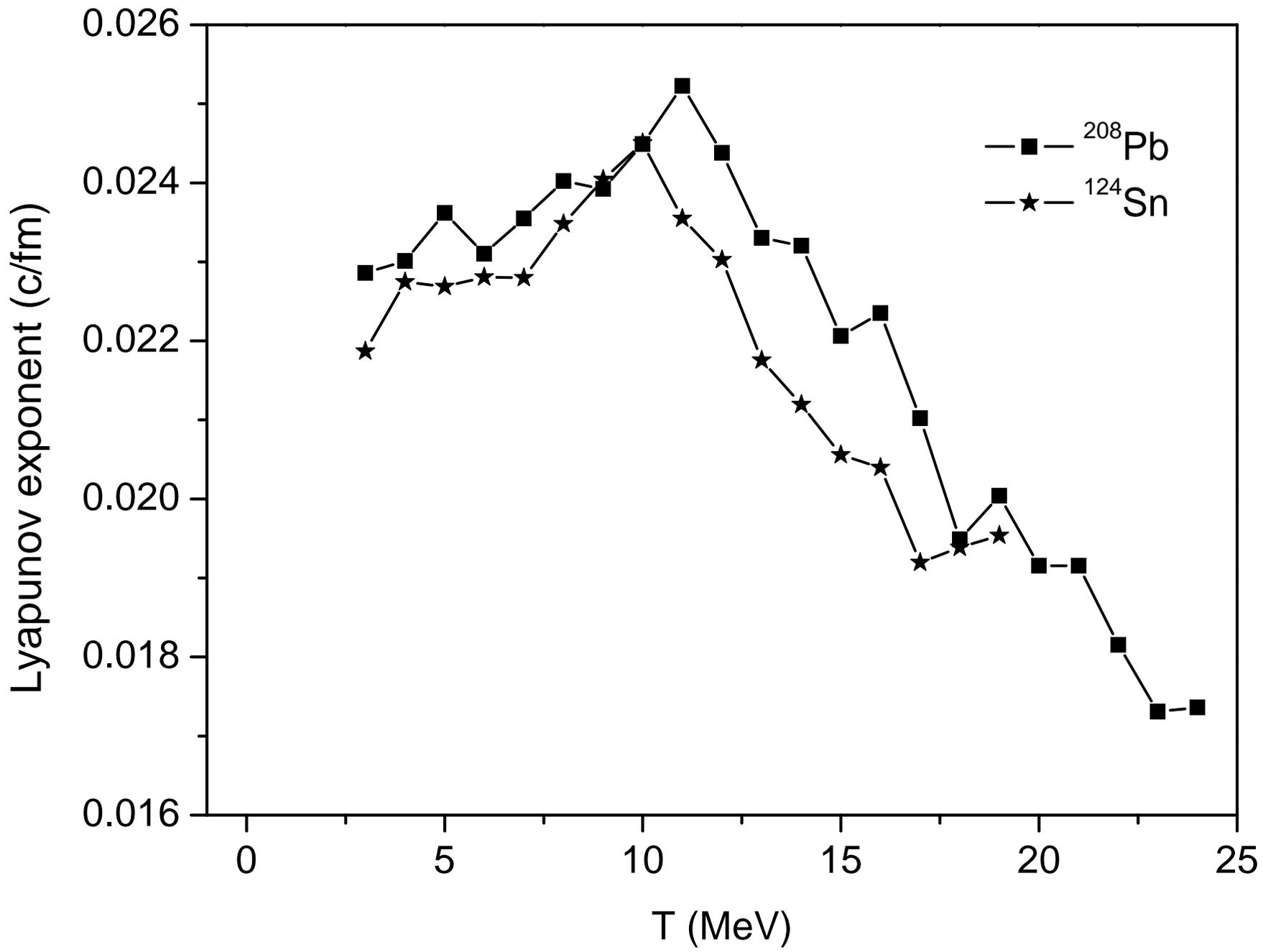
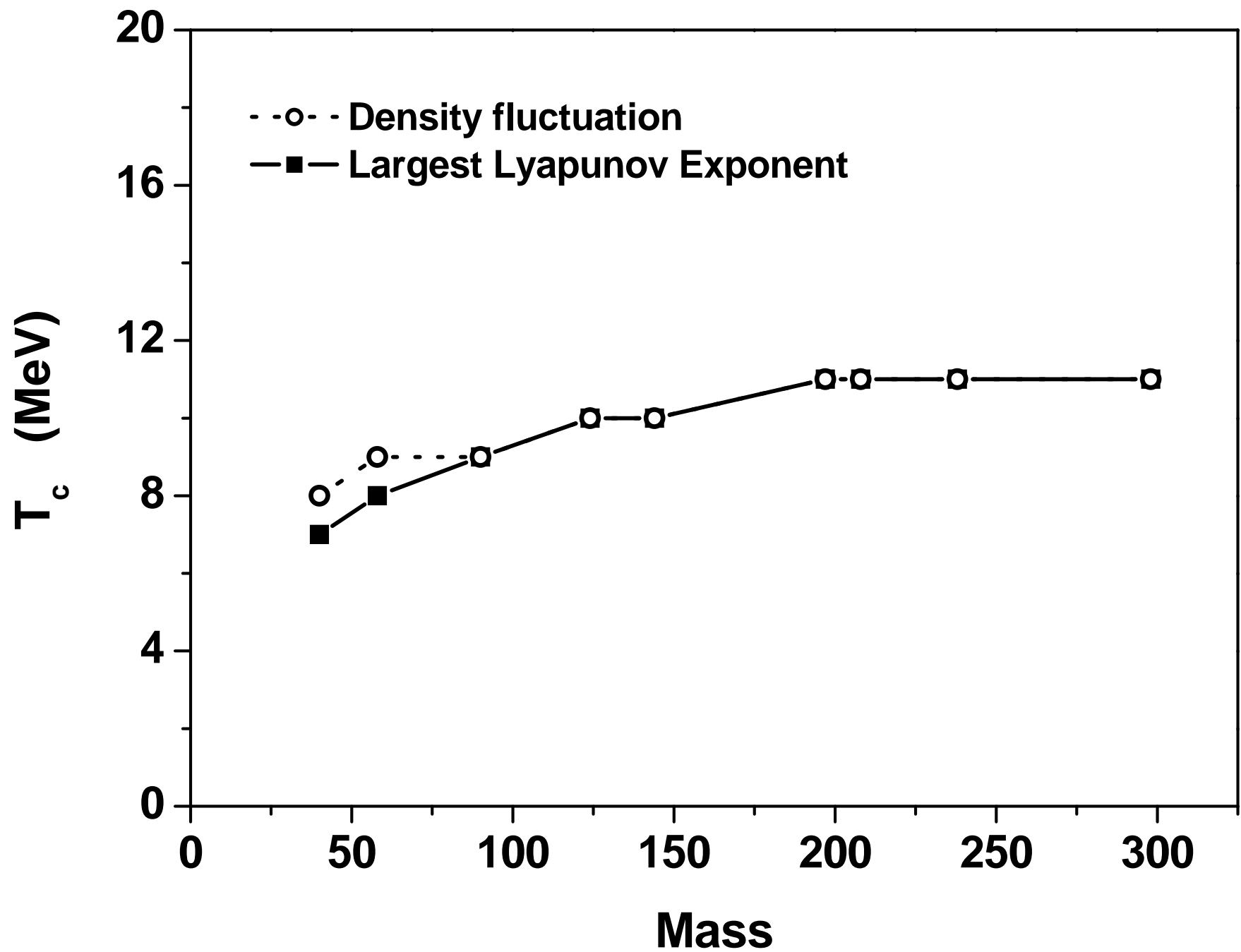
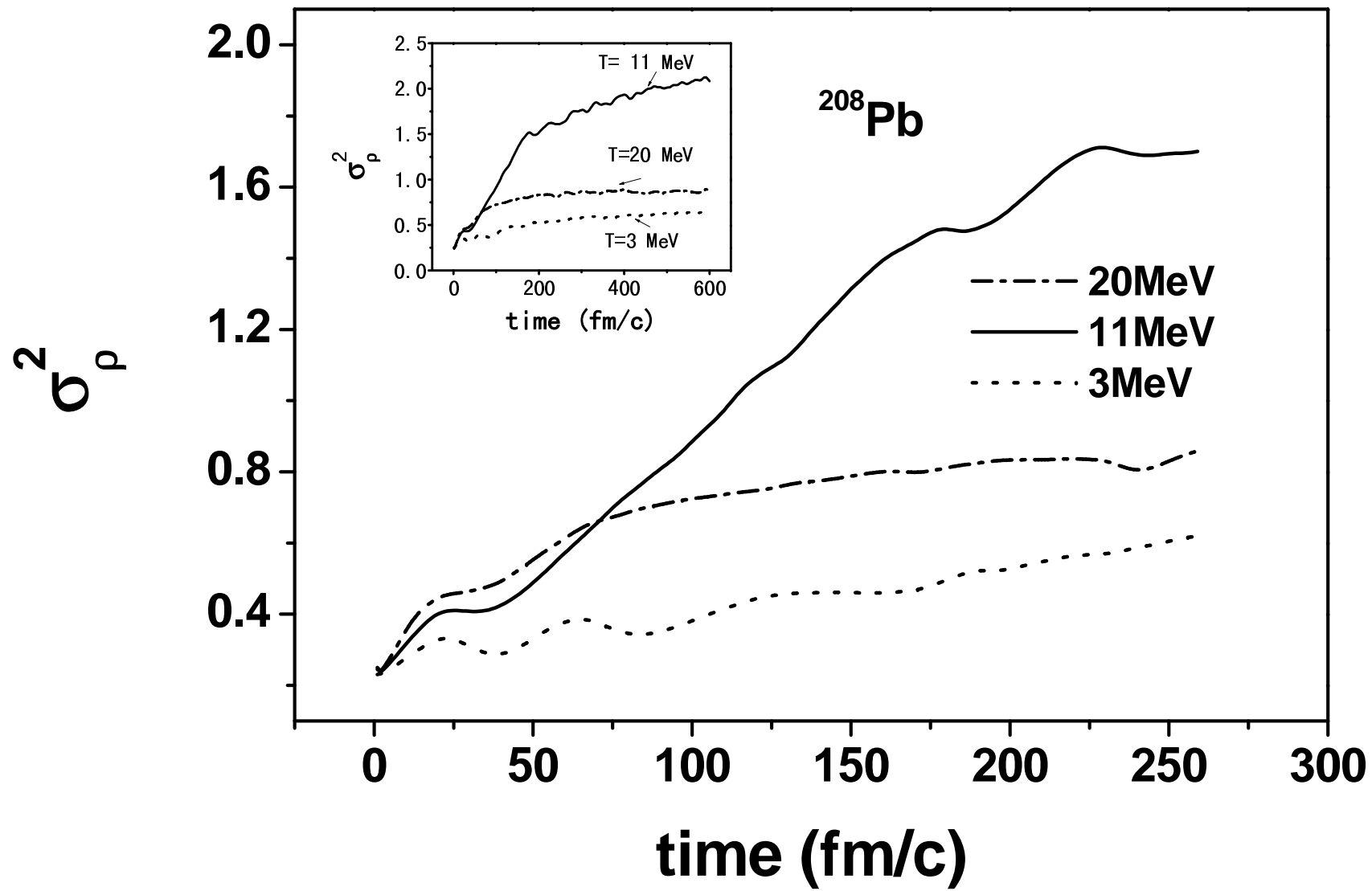
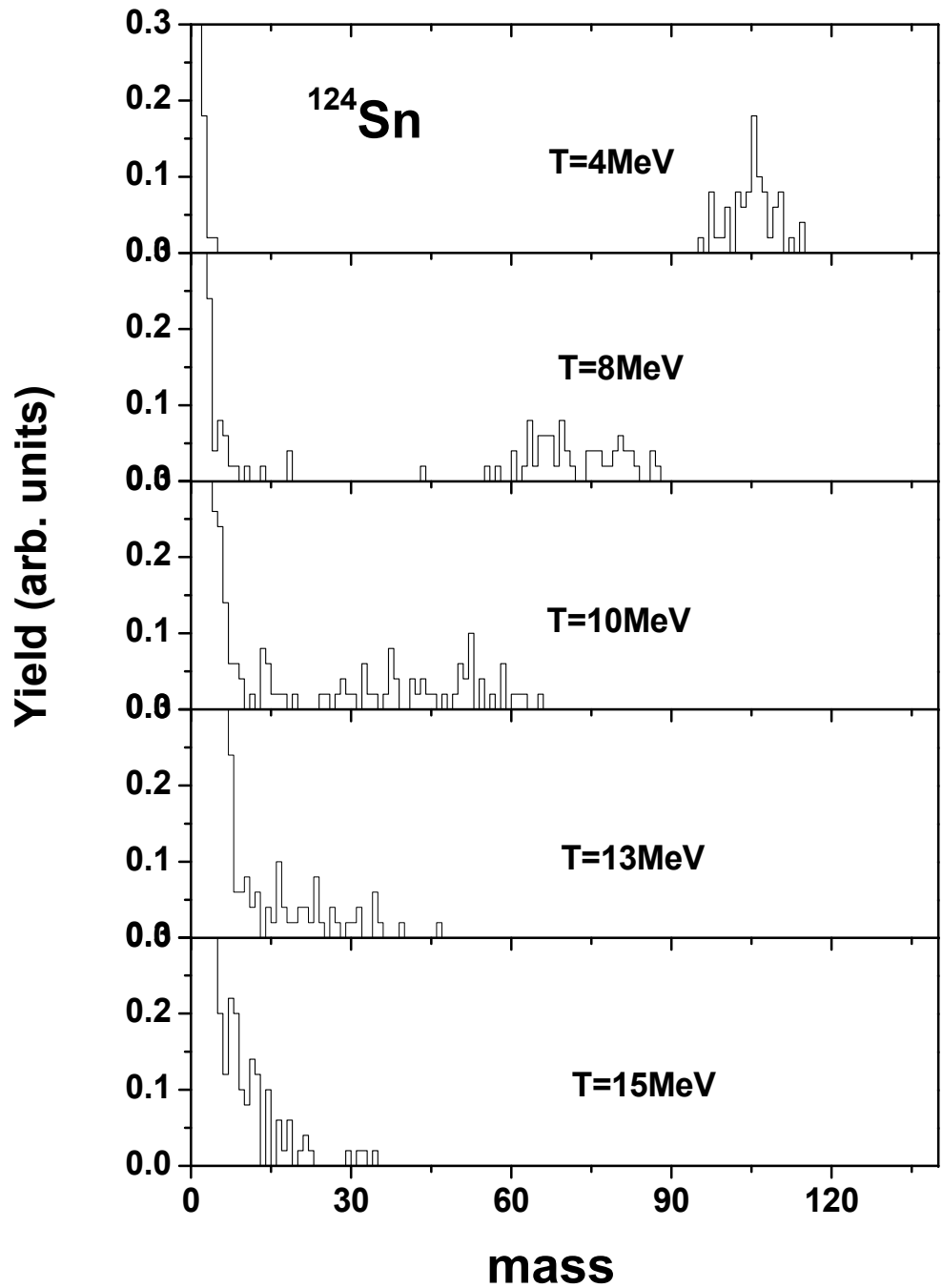
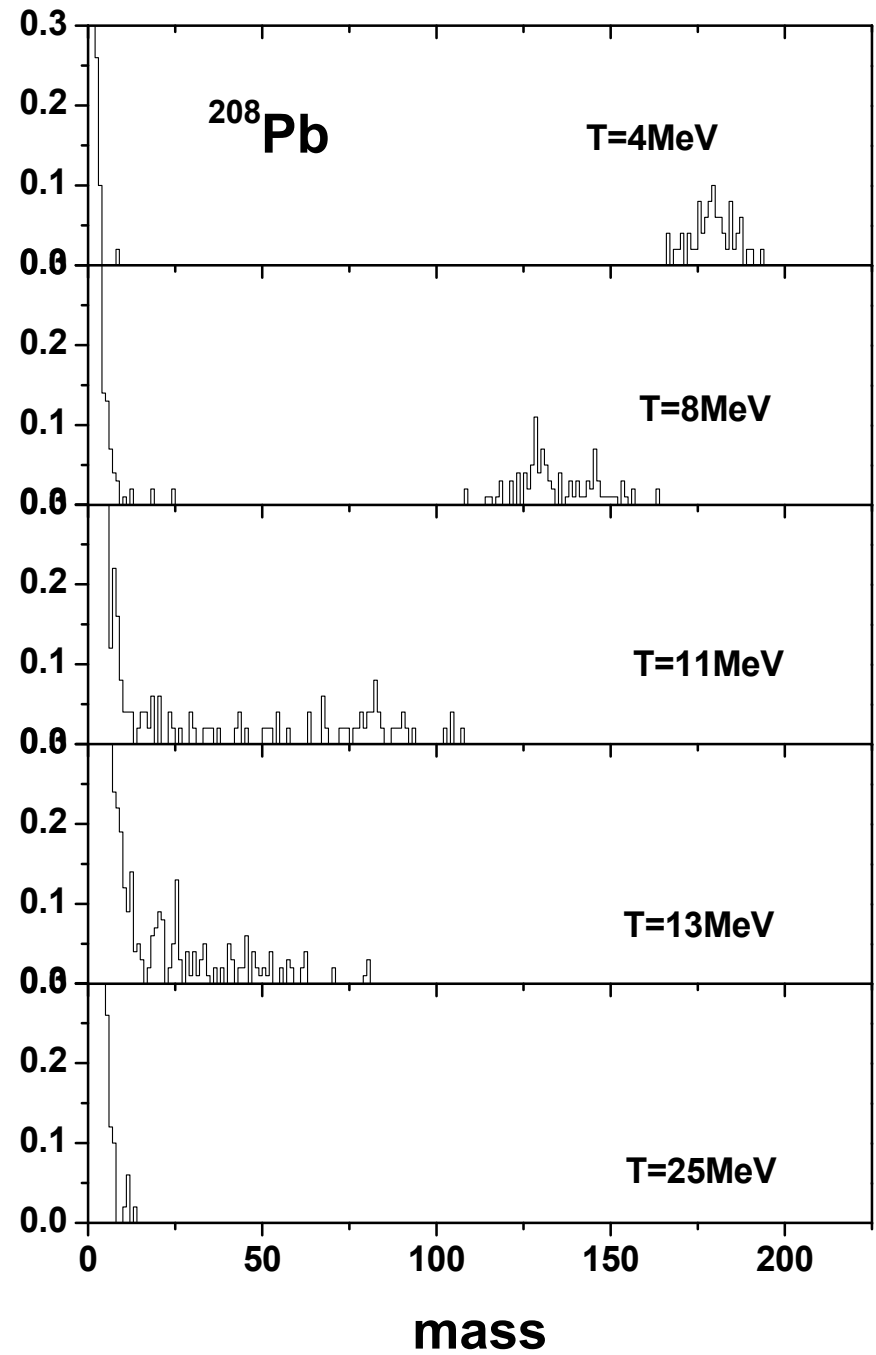
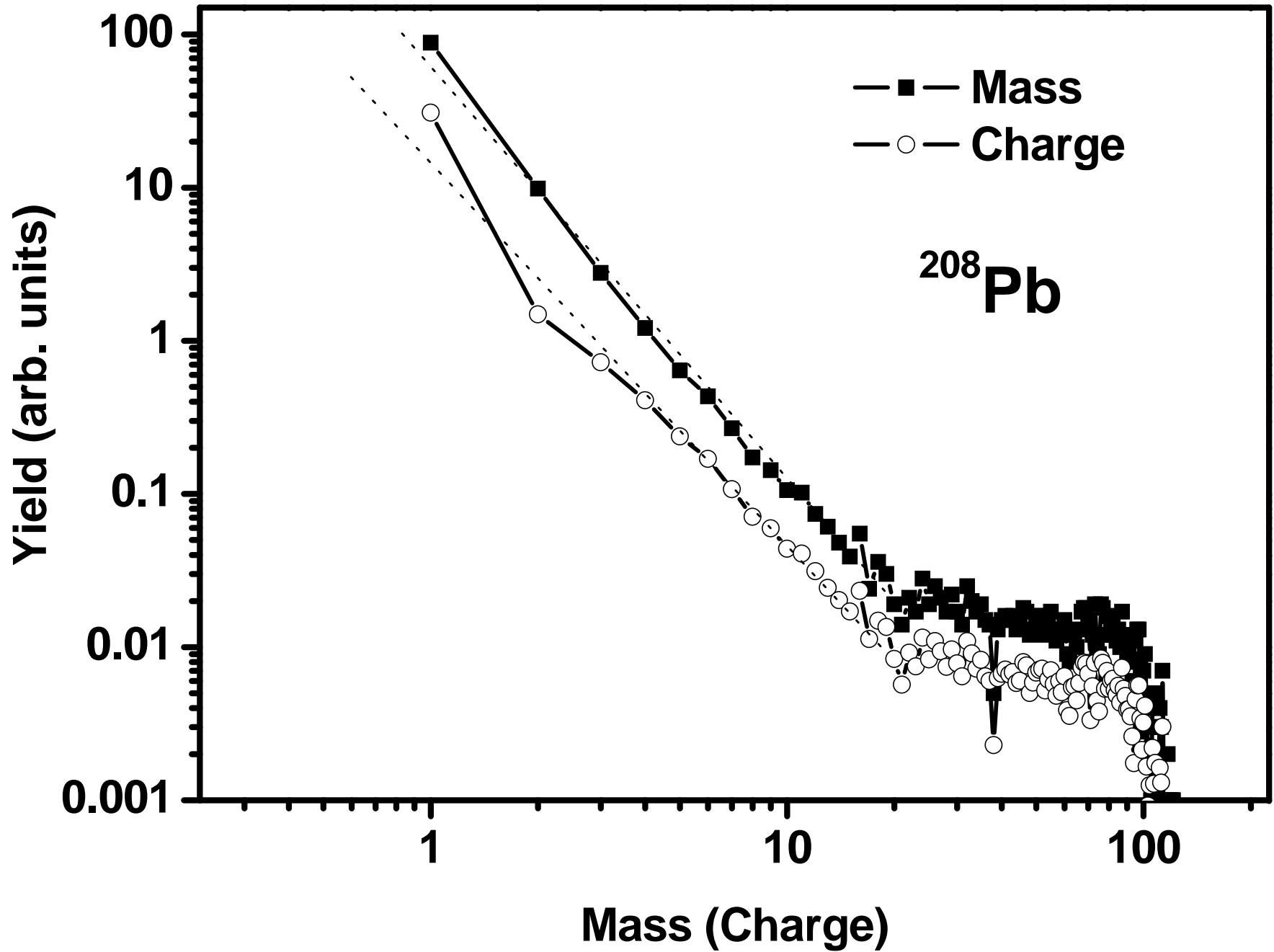
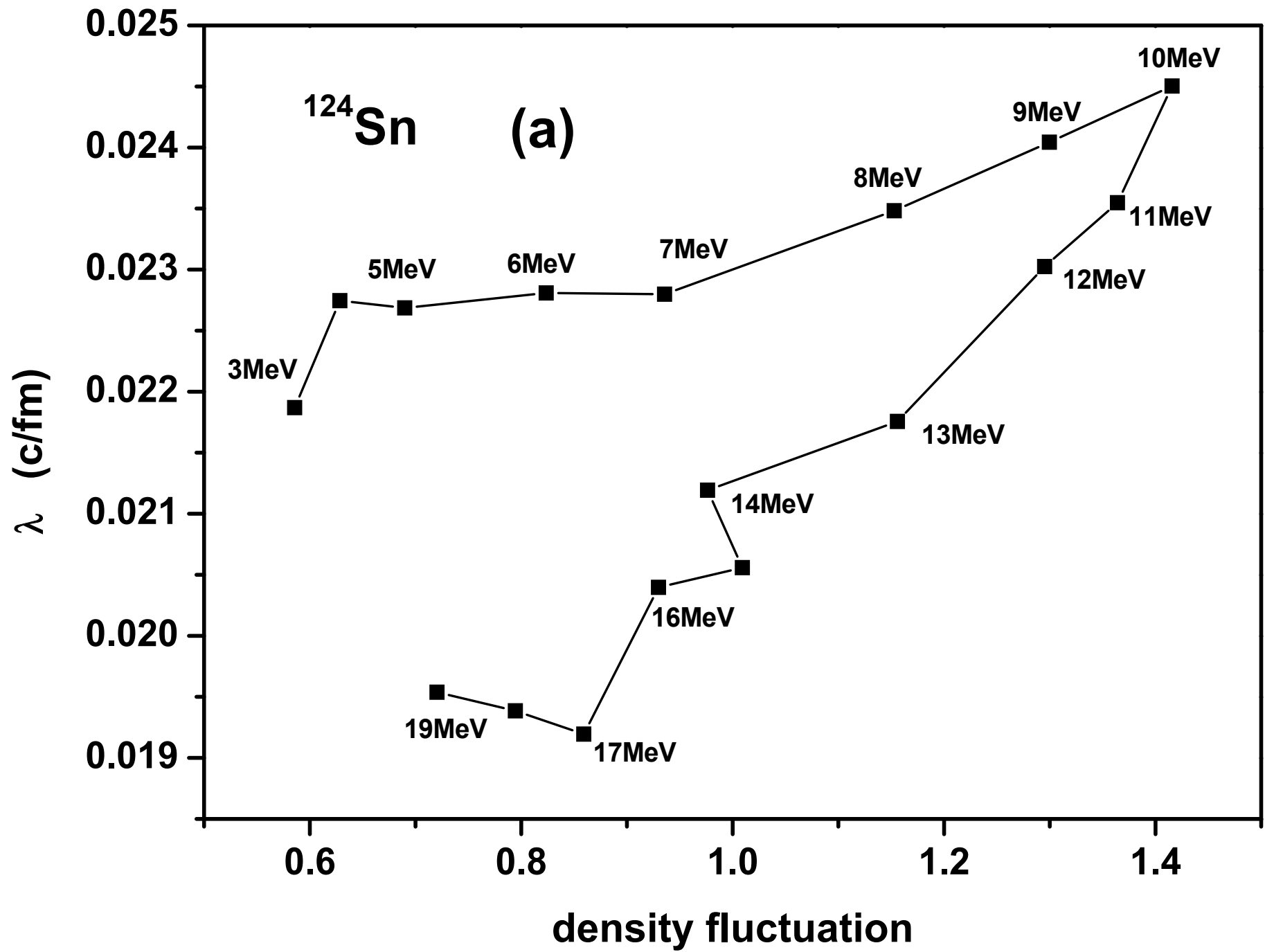
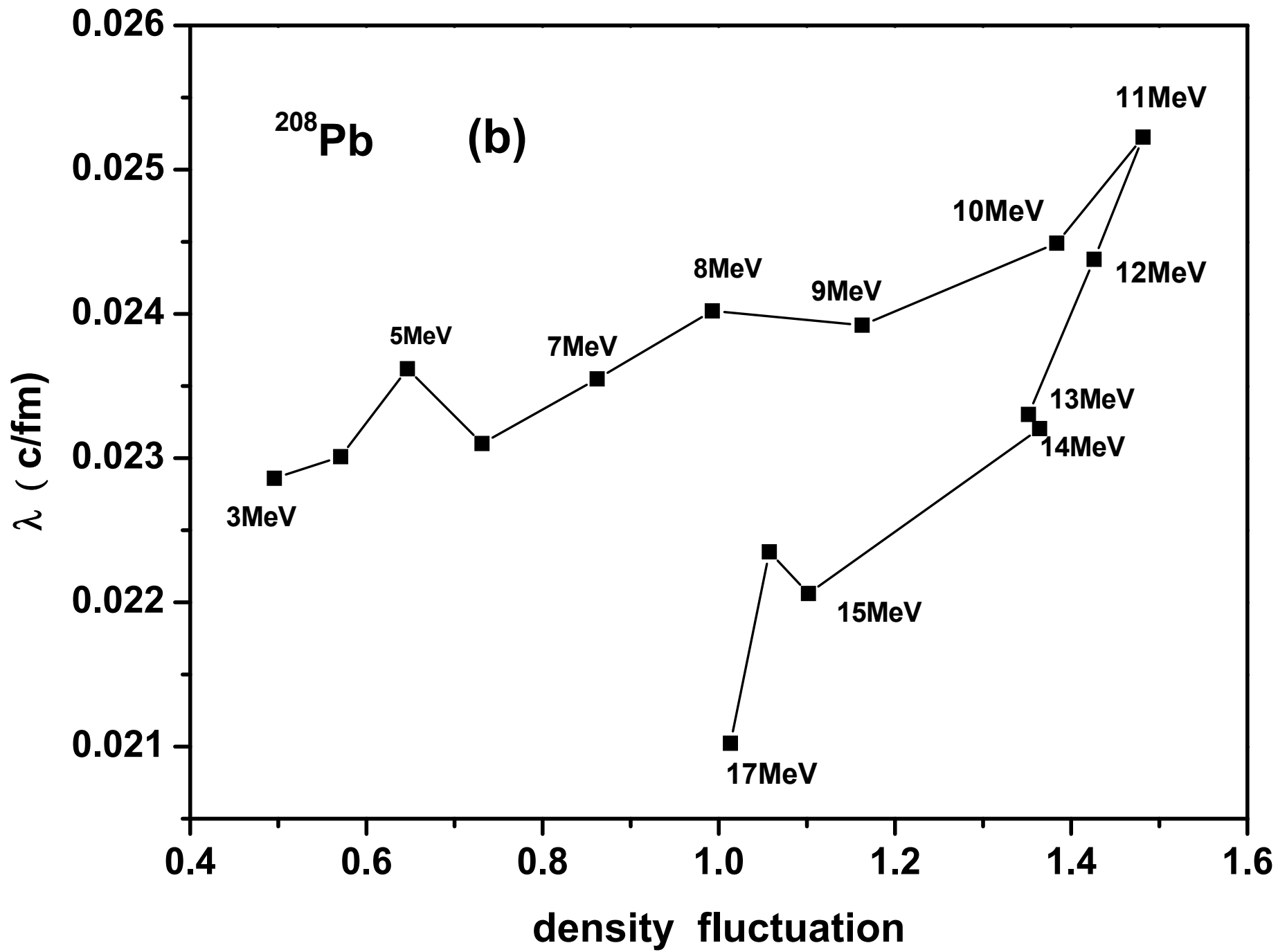
\end{document}